\documentclass[fleqn,10pt]{wlscirep}
\usepackage[utf8]{inputenc}
\usepackage[T1]{fontenc}
\title{Superfluid density and collective modes of fermion superfluid in dice lattice}

\author[1]{Yu-Rong Wu}
\author[2,3]{Xiao-Fei Zhang}
\author[4]{Chao-Fei Liu}
\author[5,6]{Wu-Ming Liu}
\author[1*]{Yi-Cai Zhang}

\affil[1]{School of Physics and Materials Science, Guangzhou University, Guangzhou 510006, People’s
Republic of China}
\affil[2]{Key Laboratory of Time and Frequency Primary Standards, National Time Service Center, Chinese Academy of Sciences, Xi'an 710600, China}
\affil[3]{School of Astronomy and Space Science, University of Chinese Academy of Sciences, Beijing 100049, China}
\affil[4]{School of Science, Jiangxi University of Technology, Ganzhou, 341000, China}
\affil[5]{Beijing National Laboratory for Condensed Matter Physics,
Institute of Physics, Chinese Academy of Sciences, Beijing 100190, China}
\affil[6]{Songshan Lake Materials Laboratory, Dongguan, Guangdong 523808, China}

\affil[*]{zhangyicai123456@163.com}


\begin{abstract}
The superfluid properties of attractive Hubbard model in dice lattice are investigated. It is found that three superfluid order parameters increase as the interaction increases. When the filling factor falls into the flat band, due to the infinite large density of states, the  resultant superfluid order parameters are proportional to interaction strength, which is in striking contrast with the exponentially small counterparts in usual superfluid (or superconductor).   When the interaction is weak, and the filling factor is  near the bottom of the lowest band (or the top of highest band), the superfluid density is determined by the effective mass of the lowest (or highest) single-particle band. When the interaction is strong and filling factor is small, the superfluid density is inversely proportional to interaction strength, which is related to effective mass of tightly bound pairs.   In the strong interaction limit and finite filling, the asymptotic behaviors of superfluid density can be captured by a parabolic function of filling factor.  Furthermore, when the filling is in flat band, the superfluid density shows a logarithmic singularity as the interaction approaches zero. In addition, there exist three undamped collective modes for strong interactions. The lowest excitation is gapless phonon, which is characterized by the total density oscillations. The two others are gapped Leggett modes, which correspond relative density fluctuations between sublattices. The collective modes are also reflected in the two-particle spectral functions by sharp peaks.  Furthermore, it is found that the two-particle spectral functions satisfy an exact sum-rule, which is directly related to the filling factor (or density of particle). The sum-rule of the spectral functions may be useful to distinguish between the hole-doped and particle-doped superfluid (superconductor) in experiments.
\end{abstract}
\begin{document}

\flushbottom
\maketitle
%
%
\thispagestyle{empty}


\section*{Introduction}

The superfluid properties in a multi-band (or multi-component) system have been attracting a great interests\cite{Peotta2015,Iskin2020,Cao2018,Wuyurong2021}.
In comparing with single-band superfluid counterpart, a lot of  novel physics, for example, existence of gapped Leggett modes \cite{Leggett1966,Iskin2005,He2016,Zhang2017,Klimin2019}, color superconductor \cite{Honerkamp2004,Rapp2007,Liu2008}, coexistence of superfluid and ferromagnetic polarizations \cite{Cherng2007,Nagy2012}, induced effective interaction \cite{Martikainen2009}, breaking of  Galilean invariance in spin-orbit coupled Bose-Einstein condensation (BEC) \cite{Zhu2012}, etc, would appear.

For a single flat band system, due to the divergence of effective mass \cite{Hazra2019}, it is believed that the flat system usually has no superfluidity, e.g., the superfluid density (which is superfluid weight up to a constant) $\rho_s=0$ .
 However, for a multi-band system, the situation may be  quite different \cite{Julku2016,Julku2020}.
 In contrast to the single-band system, the superfluid density in a multi-band system can be divided into two part contributions. One of that consists of the diagonal matrix elements of current operator (conventional part), and the other one is the off-diagonal terms (the so called geometric part \cite{Liang2017,Hu2019,Xie2020}).
Even through the conventional part may vanish in flat band limit, the geometric one may be finite.
Furthermore, it is found that under some conditions, the geometric part of superfluid density is related to the geometric quantity of energy band, namely, the geometric metric tensor of Bloch states \cite{Provost1980}.

A flat band can appear in dice ($T_3$ ) lattices,
similar as graphene, which also has the honeycomb lattice structure (see Fig. 1). In comparing with the graphene, there is an extra lattice site (per unit cell) in  the center of
 hexagon of honeycomb lattice \cite{Sutherland1986,Vidal1998,Gorbar2019}. The low-energy physics of the dice lattice are also described by the Dirac equation, but its pseudospin $S=1$ instead of $\frac{1}{2}$ \cite{Moessner2011,Xu2020}. There are two schemes to obtain this kind of lattices: one is to fabricate the trilayer structure of cubic lattice (e.g.$SrTiO_3/ SrlrO_3/ SrTiO_3$) in the (111) direction \cite{6}, and the other one is to use the cold atoms in the optical lattices \cite{20}.
The band structure of the dice lattices has a distinctive characteristic compared to the graphene, in addition to the conduction band and valence band $\epsilon_\pm (\textbf{k})$, there is a flat band $\epsilon_0 (\textbf{k})$ in the middle. For the highest (or lowest) band in the Brillouin Zone, there exists two Dirac points, which closely attach to the flat band.
Following the discovery of this new material, a lot of novel physical  phenomena, for example, Hofstadter butterfly effect \cite{Illes2016},
magneto-optical conductivity \cite{Chen2019}, Super-Klein tunneling \cite{Illes2017}, zitterbewegung effect \cite{Ghosh}, magnetism \cite{Raoux}, and
magnetoconductivity \cite{33,34}, etc., have attracted great interests.

Although the superfluid density in three band Lieb lattices has been studied \cite{Julku2016}, there are still other open questions worthwhile to investigate. For example, what roles does the effective mass play in superfluidity? How do the flat band and Dirac point affect the superfluid order parameters and superfluid density? The collective modes are also interesting subjects in such a three band system.
There usually exists several superfluid order parameters in multi-band system. The oscillations between different order parameters result in the so-called Leggett modes \cite{Leggett1966}.
The Leggett modes in condensed system have been observed experimentally in multi-band superconductor $MgB_2$, by using Raman spectroscopy \cite{Blumberg2007}, by angle-resolved photoemission spectroscopy \cite{Mou2015}.
 For three band superfluid in dice lattices,
 it is expected there may exist several Leggett modes.

In this work, we study the superfluid properties and collective modes of attractive Hubbard model in dice lattices. It is found that the flat band has important influences on the superfluid order parameters and superfluid density. In addition, there exists three collective modes, one of that is the gapless phonon mode, the other two are gapped Leggett modes.  The collective modes can be reflected by sharp peaks in two-particle spectral functions. Furthermore, an exact sum-rule of spectral functions is  derived, which may be useful to distinguish between the hole-doped and particle-doped superfluid (or superconductor) in experiments.


\begin{figure}
\begin{center}
\includegraphics[width=1.0\columnwidth]{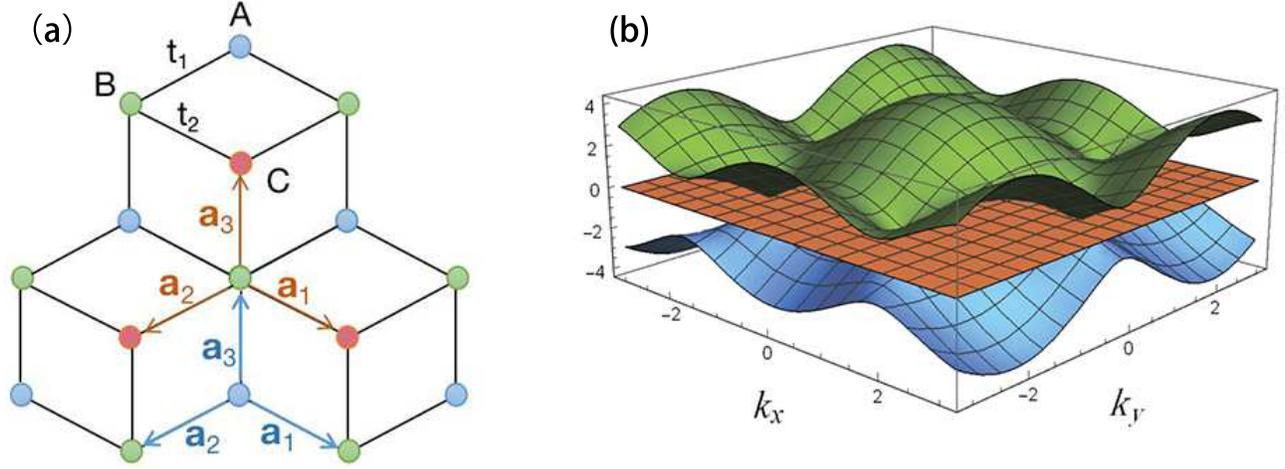}
\end{center}
\caption{ (a): The structure of dice lattices. Sublattices A, B and C are shown in blue, green and red respectively. The vectors $\textbf{a}_1=[\sqrt{3}/2,-1/2]$, $\textbf{a}_2=[-\sqrt{3}/2,-1/2]$ and $\textbf{a}_3=[0,1]$ with lattice spacing $a\equiv1$.(b): The three single-particle energy bands with $t=1$ and $\mu=0$ in Eq.(\ref{H0}).}
\label{schematic}
\end{figure}

\section*{Results}


The structure of dice lattices is depicted in Fig.1, every unit cell consists of three different lattice sites (A, B and C), in which sites A and B are located at the vertices of hexagon to form honeycomb lattice, and site C is located at the center of the hexagon. In this work, we only consider the nearest neighbor hopping and equal hopping amplitude, i.e., $t_{1}=t_{2}=t$ (see Fig. 1).
The Hubbard model in which two-component Fermi gases with spin $\sigma=\uparrow,\downarrow$ interacting through on-site attractive interaction $-U$ ($U>0$) is
 \begin{align}
    	\mathcal{H}=\sum_{\langle ij\rangle\sigma}t_{ij} \psi_{i\sigma}^\dagger \psi_{j\sigma} -\mu\sum_{i\sigma}n_{i\sigma}-U\sum_in_{i\uparrow}n_{i\downarrow},
    	\label{hamiltonian}   
\end{align}
where $\mu$ describes the chemical potential. In the following, we would use $i$ to label the unit cell. The creation operators for three sub-lattices are represented by $\psi_{iA}^\dagger$ , $\psi_{iB}^\dagger$ and $\psi_{iC}^\dagger$, respectively.  $n_{i\sigma}=\psi_{i\sigma}^\dagger \psi_{i\sigma}$ is the particle number operator. Under periodic boundary condition , using a Fourier transform, it is convenient to obtain the non-interacting Hamiltonian
\begin{align}
\mathcal{H}_0({\bf k})=\sum_{\textbf{k}} \psi^\dagger({\bf k}) h({\bf k})\psi({\bf k}),
\end{align}
  with $\psi^\dagger({\bf k})=[\psi_{A}^\dagger({\bf k}), \psi_{B}^\dagger({\bf k}), \psi_{C}^\dagger({\bf k})]$ and
\begin{align}\label{H0}
h({\bf k})=\left[\begin{array}{ccc}
-\mu &h_{12}({\bf k})  & 0\\
h_{21}({\bf k}) &-\mu &h_{23}({\bf k})\\
0 &h_{32}({\bf k}) & -\mu
  \end{array}\right]
\end{align}
where $h_{12}({\bf k})=t\sum_\delta\cos({\bf k}\cdot{\bf a}_\delta)+it\sum_\delta\sin({\bf k}\cdot{\bf a}_\delta)$, $h_{21}({\bf k})=h^{*}_{12}({\bf k})$, $h_{23}({\bf k})=h_{12}({\bf k})$, $h_{32}({\bf k})=h_{21}({\bf k})$, and ${\bf k}=(k_x,k_y)$. The two dimensional vectors ${\bf a}_\delta$ are shown in Fig.1. We note that it is a multi-band (three-band) system and there exist a flat band [$\epsilon_0(\textbf{k})=0$] in between the highest [$\epsilon_+(\textbf{k})$] and the lowest bands [$\epsilon_-(\textbf{k})=-\epsilon_+(\textbf{k})$] (see Fig. \ref{schematic}).

It is worth noting that an important role played by particle-hole symmetry in this model.  Similarly as Haldane-Hubbard model \cite{zhangyicai2017}, applying a particle-hole transformation to the system,
\begin{align}
\psi_{i\sigma}\to \epsilon_i \psi_{i\sigma}^\dagger, ~~ \psi^\dagger _{i\sigma}\to \epsilon_i \psi_{i\sigma},
\end{align}
where $\epsilon_i=1$ for the A and C sublattices, and for the B sublattices $\epsilon_i=-1$, the hopping term in Hamiltonian Eq.(\ref{hamiltonian}) is invariant under the above transformation.  After adding some constant terms, the interaction term remains invariant and chemical potential term changes a sign. At this point, we only need to consider the case where the number of particles $n>3$ ($n=3$ is half-filling, i.e. there are three particles in a unit cell). For filling $\Delta n\equiv n-3\geq0$ (particle-doped case) and $-\Delta n=3-n\leq0$ (hole-doped case), the chemical potentials satisfy a relationship
\begin{align}
\mu(-\Delta n)=-\mu(\Delta n).
\end{align}
It indicates that, at  half-filling ($n=3$), the chemical potential is exactly zero. Furthermore, we will see that all the other physical quantities, e.g., superfluid order parameters, superfluid density, etc., are also symmetrical with respect to the half-filling $n=3$ (see Fig. 2).



 

\begin{figure}
\begin{center}
\includegraphics[width=1.0\columnwidth]{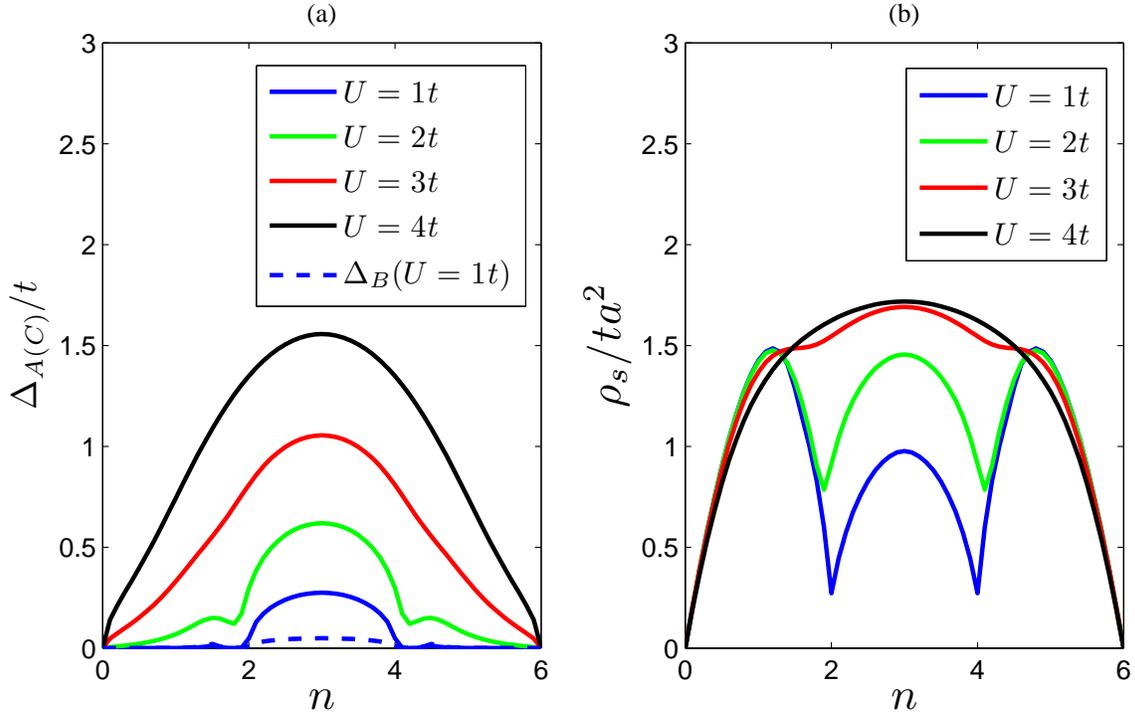}
\end{center}
\caption{  The superfluid order parameters for sublattice A(C) and superfluid densities as functions of filling factor $n$. (a):  The superfluid order parameters for interaction $U=1$, $U=2$, $U=3$ and $U=4$ are shown in blue, green, red and black solid lines, respectively. The blue dashed line is the order parameter $\Delta_B$ when $U=1$. (b): The superfluid densities for interaction $U=1$, $U=2$, $U=3$ and $U=4$ are also shown in blue, green, red and black solid lines, respectively.   }
\label{pdm01}
\end{figure}

The order parameters ($\Delta_i=-U\langle \psi_{i\downarrow}\psi_{i\uparrow}\rangle$) can be calculated with mean field theory (see\textbf{ Metholds}).
 The evolutions of the superfluid order parameters are depicted in panel (a) of Fig. 2. It is found that $\Delta_i$ is always real. The order parameters are symmetrical with respect to the half-filling due to the particle-hole symmetry.  The Hamiltonian is subject to the inversion symmetry, $\Delta_A$ is the same as $\Delta_C$ but differ from $\Delta B$. As one increases the interaction $U$, $\Delta_ i$  always increases. The order parameters can increase without limit as long as the interaction is strong enough. For example, when filling factor is half-filling ($n=3$), $\Delta_i\simeq U/2$ as $U\rightarrow \infty$, which is a half of the energy of two particle bound states \cite{Heiselberg2012}.

When the filling factor approaches zero, the order parameters decrease to zero. The fully occupied case ($n=6$) is equivalent to completely empty ($n=0$) case due to the particle-hole symmetry. With the increasing of filling factor $n$, the $\Delta_i$ gets larger. However, when the filling factor is in between the lowest band and the middle flat band, the order parameter gets a dip due to the small density of states near the Dirac points [especially for weakly interacting cases in panel (a) of Fig.2]. The suppressions of order parameters are  also found in Lieb lattices \cite{Iglovikov2014}. When the filling factor lies in the middle region (e.g., $n\simeq3$), the order parameter reaches its maximum  due to enhancement of density of states in flat band.

The interaction effects amplified by band flatness can give rise to Wigner-crystal state \cite{Wu2008} and ferromagnetic transition \cite{Zhang2010}.
Similarly, the behaviors of superfluid order parameters are also affected significantly by the flat band \cite{Kopnin2011}. For example, when the  filling factor is in the middle of the flat band (e.g., half filling $n=3$), the infinite large  density of states results in order parameter $\Delta_A=\Delta_C\simeq U/4$ as the interaction approaches zero, which is very different from that in the usual Bardeen-Cooper-Schrieffer (BCS) superconductor (or superfluid), where pairing gap $\Delta$ is exponentially small when interaction $U\rightarrow0$.
In addition, for half-filling $n=3$, it is found that the order parameter for sublattice B is much smaller than $\Delta_{A(C)}$ when the interaction is small ($U/t\ll1$), i.e., $\Delta_B/\Delta_{A(C)}\ll1$ (see Fig.\ref{pdm01}). \textbf{This is because the weights of sublattices A and C in flat band wave function is dominant over that of sublattice B.}

\subsection*{Superfluid density}
The superfluid density can be calculated with phase twist method \cite{Fisher1973,Liang2017}.  Assuming the superfluid order parameters undergo a phase variation, e.g, $\triangle_i\rightarrow \Delta_ie^{2i\textbf{q}\cdot\textbf{r}_i}$,
the superfluid density (particle number per unit cell) tensor $\rho_{sij}$ can be written as
 \begin{eqnarray}
 \rho_{sij}= \frac{\partial^2 \Omega(\textbf{q})}{\partial q_i \partial q_i}|_{q\rightarrow0},
 \label{phasetwist}
\end{eqnarray}
where $\Omega$ is thermodynamical potential (per unit cell) in grand canonical ensemble. Similarly as the Haldane model \cite{Liang2017}, due to the $C_3$ symmetry of honeycomb lattices \cite{Julku2020,Zhangyicai2020}, the superfluid density tensor can be simplified into a scalar, i.e., $\rho_{sij}=diag\{\rho_s,\rho_s\}$.

The superfluid density $\rho_s$ is plotted as a function of filling factor for different interactions in panel (b) of Fig. 2. First of all, we see that the superfluid density is also symmetrical with respect to the half-filling due to the particle-hole symmetry. When the filling factor is completely empty ($n\rightarrow0$) [or fully occupied case ($n\rightarrow6$)], the superfluid density vanishes linearly.
  For the small filling factor  ($n\ll 1$) and weak interaction ($U\ll t$) case (BCS-limit \cite{Pieri}), the superfluid density can be written as
 \begin{align}
   \rho_s(U/t\ll 1)=n/m^*=3tn/(\sqrt{2})\simeq2.12nt,
    \end{align}
 where $m^*=\sqrt{2}/(3t)$ is the effective mass near the bottom of lowest band, which is determined by the single-particle energy band structure,  and does not depend on interaction strength $U$. The superfluid density is proportional to  filling factor (or particle density) and inversely proportional to the effective mass of energy band \cite{Hazra2019}, which is very similar to that of Bose-Einstein condensate with spin-orbital coupling \cite{zhangyicai2016}.

 Differently from order parameters, when interaction $U/t\gg 1$, the superfluid density remains finite. In such case, the superfluid consists of tightly bound pairs. Two bound pairs could not occupy a same lattice site due to the constraint of exclusive principle of fermions. So the pairs can be viewed as hard-core bosons, which is BEC limit of fermion superfluid.
When the number of the effective hard-core bosons is small $n\ll 1$ (dilute gas limit), the superfluid density is determined by the effective mass of tightly bound pairs, i.e.,
\begin{align}
   \rho_s(U/t\gg 1)=n'/m'^*=8t^2n/U,\label{effectivemass}
    \end{align}
where $n'=n/2$ is filling factor of effective hard core bosons and $m'^*=U/(16t^2)$ is its effective mass, which is determined by the dispersion of two-particle bound state energy (Note $m'^*\neq2m^*$ due to the coupling of between the motions of center of mass and relative motion of two particles in lattice \cite{Zhangyicai2013}). The above equation indicates that when the interaction gets stronger (in strong interaction limit), the effective mass becomes larger, the resultant superfluid density gets smaller.
For a moderate interaction strength (e.g., $U/t>5$), and small filling ($n\ll 1$), it is expected that the superfluid density should satisfy  $8t^2n/U<\rho_s<2.12nt$.

\begin{figure}
\begin{center}
\includegraphics[width=0.900\columnwidth]{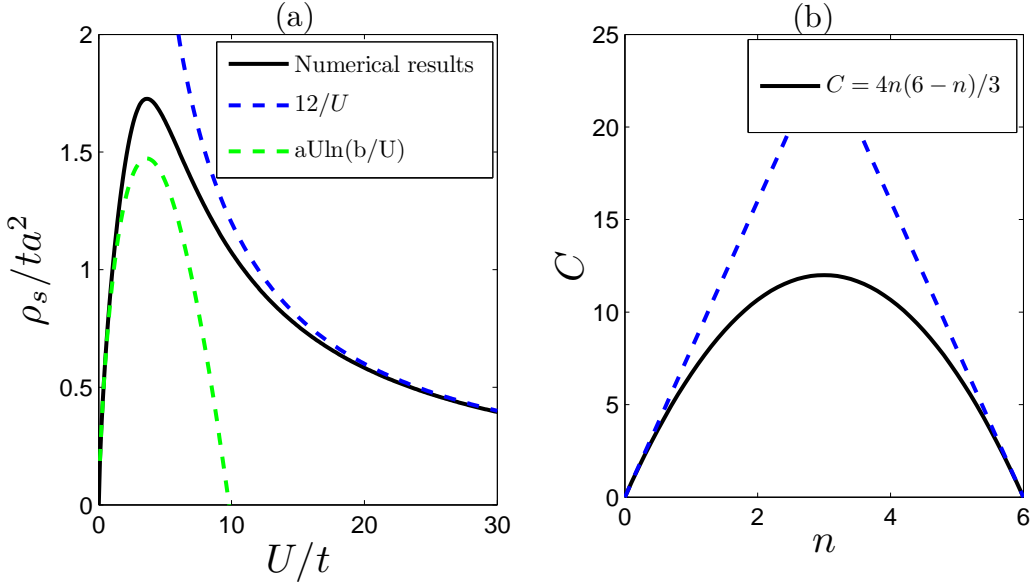}
\end{center}
\caption{ The asymptotic behaviors of superfluid densities. (a):  The numerical results of superfluid density are plotted in black solid line (filling factor $n=3$).
The two dashed lines correspond two asymptotic limits of superfluid density, which are given by $\rho_s\simeq 12/U$ for $U\rightarrow\infty$) and $aUln(b/U)$ with $a=0.4077$ and $b=9.822$ for $U\rightarrow0$, respectively.
 (b): The blue dashed line is the estimated up bound  $8n$ (filling factor $0\leq n\leq3$), which is determined by effective mass of tightly bound pairs [$m^{'*}=U/(16t^2)$]. Using the particle-hole symmetry, the corresponding up bound for $3\leq n\leq6$ also depicted here. The solid line is the parabolic function Eq.(\ref{hardcore}). }
\end{figure}

 In addition, differently from the order parameters,  the evolution trends of  superfluid density with the interactions can bedifferent for different regimes of filling factors. For example, when the filling is small, superfluid density diminishes with the increasing of interaction, which is opposite to that of order parameters [see panel (b) of Fig. 2]. Only when the filling is sufficiently large (for example, $n\simeq3$), the superfluid density grows up as the interaction increases (within a certain range of interaction, see the following discussions and the panel (a) of Fig. 3).
 Due to the vanishingly small density of states near the Dirac points, the superfluid density also develops a dip for weak interacting cases. Similarly as Lieb lattices \cite{Julku2016}, a triple dome structure appears in the superfluid density, especially for weakly interacting cases [see panel (b) of Fig. 2].

    For intermediate filling factor (e.g, $n=3$ in flat band), it is found that the superfluid density approaches zero as $U\rightarrow0$.
     Once the interaction is turned on, the superfluid density grows rapidly [see panel (a) of Fig. 3]. Specially here,  the non-vanishing superfluid density for weak interactions mainly arises from the contributions of off-diagonal matrix elements of current operator (the geometric part).  In the isolated flat band limit \cite{Julku2016,Xie2020} or under the assumption of uniform pairing \cite{Peotta2015,Liang2017},  the superfluid density in flat band can be related to geometric metric tensor of Bloch sates.
     However, here we note that the superfluid density approaches zero according to the law of
      \begin{align}
      \rho_s=aUln[b/U],
      \end{align}
       as $U\rightarrow 0$ with two constants $a\simeq0.4077$ and $b=9.822$ for half filling $n=3$ [see panel (a) in Fig.3], which is different from the linear dependence in isolated flat band limit \cite{Julku2016}. As $U\rightarrow0$, the appearance of the logarithmic singularity ($UlnU$) is attributed to the existences of the gapless Dirac points and vanishingly small pairing gap ($\Delta\propto U\rightarrow0$).  This is because when  filling factor is in the middle of flat band (half-filling of $n=3$) and $U\rightarrow0$ (or $\Delta\equiv\Delta_A=\Delta_C\simeq U/4 \rightarrow 0$), it is found that the superfluid density mainly arises from matrix elements of current operator between the flat band and other bands near the Dirac points (the off-diagonal part). Furthermore, the contribution near the Dirac points can be approximated by 
       \begin{align}\rho_s\sim A\Delta \int_{0}^{B} \frac{ qdq}{\sqrt{q^2+\Delta^2}(\sqrt{q^2+\Delta^2}+3\Delta)}=A\Delta ln\frac{\sqrt{B^2+\Delta^2}+3\Delta}{4\Delta}\simeq a U ln(b/U),
       \end{align}  where $A$, $B$, $a$ and $b$ are four constants. \textbf{A similar logarithmic singularity of superfluid density also appear in the Mielke checkerboard lattices \cite{Iskin2019}.}

        With the increasing of interaction, the superfluid density reaches its maximum values at $U\simeq3.7t$ [see panel (a) in Fig.3]. After passing over that point, it decreases.
  For strong interaction limit and finite $n$, the trend of evolutions of superfluid density can be also understood qualitatively by effective mass of tightly bound pairs. Considering the fermion nature of tightly bound pairs,  it is expected that the superfluid density should be smaller than that determined by $m'^*$, namely, $\rho_s< 8t^2n/U$  for BEC limit ($U\gg t$).
   For example, when $n=3$, it is found that the superfluid density is also inversely proportional to interaction, i.e., $\rho_s\rightarrow C/U$ as $U\rightarrow\infty$ [see panel (a) of Fig.3]. The numerical results give the constant $C\simeq12$ for half-filling case ($n=3$), which is almost a half of the estimated up bound [$8t^2n=24$ (t=1)].   Nevertheless, the up bound would provide much more precise estimation for superfluid density when filling factor $n$ is small [see panel (b) of Fig.3].
For a general filling $n$, it is found that the constant $C$ can be given by a parabolic function (see the \textbf{Methods}), namely
   \begin{align}
   C=\frac{4t^2}{3}n(6-n).
   \label{hardcore}
    \end{align}
Consequently, the superfluid density 
\begin{align}
   \rho_s=\frac{4t^2}{3U}n(6-n).
   \label{hardcore1}
    \end{align}
The asymptotic results  Eqs. (\ref{hardcore}) and (\ref{hardcore1}) in strong interaction limit are very similar to the behaviors of the superfluid density of hard-core bosons in lattices \cite{Onogi1994,Micnas1995}.

\subsection*{Collective modes}
In this section, we would present Gaussian fluctuation method \cite{Engelbrecht1997,Hu2006,Diener2008} to calculate the collective modes and spectral functions  \cite{zhangyicai2017}.
The order parameters can be decomposed as mean-field part and fluctuation, i.e., $\Delta_i=\Delta_i+\delta \Delta_i$ with $i=1,2,...m$, and $m$ is the number of order parameters. The partition function can be written as
\begin{eqnarray}
 && Z\approx e^{-S_0}\int D \eta^{\dag}_q D\eta_q e^{-\delta S},
\end{eqnarray}
where $S_0$ is the mean-field contribution and  the Gaussian fluctuation part (see the \textbf{Methods})
\begin{eqnarray}
 & \delta S=\frac{1}{2}\sum_{\textbf{q},n}\eta^{\dag}_q M(q)\eta_q=\sum_{\textbf{q},n>0}\eta^{\dag}_q M(q)\eta_q,
\end{eqnarray}
with pairing fluctuation fields $\eta^{\dag}_q=[\Delta^{*}_{1q},\Delta^{*}_{2q},...,\Delta^{*}_{mq},\Delta_{1,-q},\Delta_{2,-q},\Delta_{m,-q}]$ and $q=(\textbf{q},i\omega_n)$, $\omega_{n}=2n\pi/\beta$ ($n\in Z$) is Matsubara frequency, $\beta=1/T$ is inverse temperature. The
 fluctuation matrix $M$ is a $2m\times2m$  matrix,
\begin{eqnarray}
 &&M_{ij}(\textbf{q},i\omega_n)=\frac{1}{\beta}\sum_{\textbf{k},n'}G^{0}_{ij}(k+q)G^{0}_{j+m,i+m}(k)+\frac{\delta_{ij}}{U}, \qquad \qquad  \qquad\qquad (1\leq i,j\leq m)\notag\\
 &&M_{ij}(\textbf{q},i\omega_n)=\frac{1}{\beta}\sum_{\textbf{k},n'}G^{0}_{ij}(k+q)G^{0}_{j-m,i+m}(k), \qquad \qquad \qquad \qquad (1\leq i\leq m \;\&\; m+1\leq j\leq 2m)\notag\\
&&M_{ij}(\textbf{q},i\omega_n)=\frac{1}{\beta}\sum_{\textbf{k},n'}G^{0}_{ij}(k+q)G^{0}_{j+m,i-m}(k), \qquad  \qquad \qquad \qquad(m+1\leq i\leq 2m \; \& \; 1\leq j\leq m)\notag\\
 &&M_{ij}(\textbf{q},i\omega_n)=\frac{1}{\beta}\sum_{\textbf{k},n'}G^{0}_{ij}(k+q)G^{0}_{j-m,i-m}(k)+\frac{\delta_{ij}}{U}, \qquad \qquad \qquad \qquad(m+1\leq i,j\leq 2m ),
\end{eqnarray}
where
\begin{eqnarray}
 G^{0}_{ij}(k)=([i\omega_{n'}-H_{\rm BdG}(\textbf{k})]^{-1})_{ij},
\end{eqnarray}
is matrix element of Nambu-Gorkov Green function. Here $k=(\textbf{k},i\omega_{n'})$ and $\omega_{n'}=(2n'+1)\pi/\beta$ [$n'\in Z$].

 The collective modes are given by zeros of determinant $Det|M(\textbf{q},i\omega_n\rightarrow \omega+i0^+)|=0$. As $q\rightarrow0$, the gapless collective mode is the Anderson-Bogoliubov phonon, which characterizes the density oscillations of superfluid. The gapped ones are Leggett modes, which correspond relative oscillations of densities of different sublattices \cite{Leggett1966,Iskin2005,He2016,Zhang2017,zhangyicai2017}.

The two-particle normal (anomalous) Green function matrix elements are defined as \cite{Zhangyicai2020}
\begin{align}
& G_{II,ij}(\textbf{q},i\omega_n)=\sum_n[\frac{\langle0|\Delta_{i \textbf{q}}|n\rangle\langle n|\Delta^{\dag}_{j \textbf{q}}|0\rangle}{i\omega_n-\omega_{n0}}-\frac{\langle0|\Delta^{\dag}_{j \textbf{q}}|n\rangle\langle n|\Delta_{i \textbf{q}}|0\rangle}{i\omega_n+\omega_{n0}}],\notag\\
& F_{II,ij}(\textbf{q},i\omega_n)=\sum_n[\frac{\langle0|\Delta_{j,\textbf{q}}|n\rangle\langle n|\Delta_{i,-\textbf{q}}|0\rangle}{i\omega_n-\omega_{n0}}-\frac{\langle0|\Delta_{i,-\textbf{q}}|n\rangle\langle n|\Delta_{j,\textbf{q}}|0\rangle}{i\omega_n+\omega_{n0}}],
 \label{definition}
\end{align}
where $\omega_{n0}=E_n-E_0$, $E_n$ and $|n\rangle$  are eigenvalues and eigenstates of Hamiltonian, respectively. $\Delta_{i\textbf{q}}$ is pairing fluctuation operator of superfluid order parameters [see Eq.(\ref{order}) in the next section].
With the action $\delta S$ (Gaussian weight), the  correlation functions (and the matrix elements of two-particle Green functions)  can be calculated \cite{Zhangyicai2020}, i.e.,
 \begin{eqnarray}
 &&-G_{II,ij}(q)=\langle \Delta^{*}_{jq} \Delta_{iq}\rangle=(M^{-1})_{i,j},\notag\\
 &&-F_{II,ij}(q)=\langle \Delta_{j,-q} \Delta_{iq} \rangle=(M^{-1})_{i,m+j},\notag\\
&& -F^{*}_{II,ji}(q)=\langle \Delta^{*}_{jq}\Delta^{*}_{i,-q} \rangle=(M^{-1})_{m+i,m+j},\notag\\
 && -G_{II,ji}(-q)=\langle  \Delta_{j,-q}\Delta^{*}_{i,-q}\rangle=(M^{-1})_{m+i,m+j}.
\end{eqnarray}

The two-particle spectral functions can be expressed in terms of the Green functions as \cite{Samanta}
\begin{align}\label{spectral}
  &A_{ij}(\textbf{q},\omega)=-\frac{1}{\pi}Im [G_{II,ij}(\textbf{q},i\omega_n\rightarrow\omega+i0^+)],\notag\\
   & B_{ij}(\textbf{q},\omega)=-\frac{1}{\pi}Im [F_{II,ij}(\textbf{q},i\omega_n\rightarrow\omega+i0^+)],
\end{align}
where $0^+$ denotes a positive infinitesimal number.
In addition, in the presence of inversion symmetry,
 the superfluid density can be also related to the pairing fluctuation matrix $M$ \cite{Zhangyicai2020}
 \begin{eqnarray}
 \rho_s\!\!=\!\!{\rm lim_{q\rightarrow0}}\frac{4}{q^2}(\Delta^{t*},\Delta^t).\!\!\left( \!\!\!                
 \begin{array}{cccc}   
  I  & 0 \\  
   0&  -I\\  
\end{array}\!\!\!\right)\!\!.M(q).\!\!\left( \!\!\!                
 \begin{array}{cccc}   
  I  & 0 \\  
   0&  -I\\  
\end{array}\!\!\!\right)\!\!.\!\!\left(\!\!\!
 \begin{array}{cccc}   
   \Delta \\  
   \Delta^*\\  
\end{array}\!\!\!\right).
\label{spdensity3}
\end{eqnarray}
where $(I)_{m\times m}$ is a $m\times m$ identity matrix, $\Delta=(\Delta_1,\Delta_2,...,\Delta_m )^t$, and $(...)^t$ denotes matrix transpose. In dice lattice case, the number of order parameters $m=3$ and we identify $\Delta_1=\Delta_A$, $\Delta_2=\Delta_B$ and $\Delta_3=\Delta_C$.

Fig. 4 and 5 show the collective modes with the increasing of wave vector $q$. It is found that there exist three undamped collective modes for strong interactions [see panels (a) and (b) of Fig. 5]. The lowest one is the gapless phonon as $q\rightarrow0$, which corresponds to total density oscillation. The upper two gapped excitations are the Leggett modes, which corresponds to the relative density oscillations between sublattices.
As the interaction strength decreases, the bottom of two-particle continuum lowers, the region of the existence of collective modes shrinks.  The collective modes merge into the continuum and becomes damped [see panels (a) and (b) of Fig. 4]. As a result of that, the collective modes are not well defined. Only when the interaction is strong enough, the undamped Leggett modes survive (see Fig. 5).
The collective modes also reflected in the two-particle spectral functions as shown in Fig. 4 and Fig. 5. The sharp peaks in spectral functions correspond the collective modes [see panels (c) and (d) in Fig. 4 and 5].
\textbf{The phonon is in-phase oscillation of three order parameters, while the other two Leggett modes are out-phase ones.}

\begin{figure}
\begin{center}
\includegraphics[width=12.00cm]{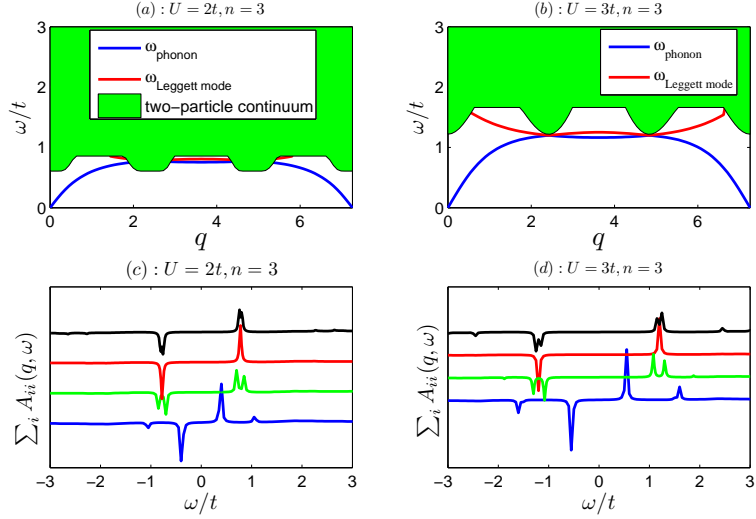}
\end{center}
\caption{  Collective modes [panel (a) and (b)] in x-direction $\textbf{q}=q \hat{x}$ and two-particle spectral functions [panel (c) and (d)]. (a): The phonon (Leggett) mode is plotted in blue (red) line. The green shaded region is the two-particle continuous spectra. The interaction $U=2$, filling factor $n=3$ in panel (a).
(b): The phonon mode and Leggett mode with $U=3;n=3$.  (c): Spectral functions with $U=2;n=3$.  (d): Spectral functions with $U=3;n=3$. The spectral functions of $q=0.5, 1.5, 2.5$ and $3.5$ are plotted in blue, green, red and black lines, respectively. The sharp peaks in spectral functions correspond the collective modes. In order to get the spectral functions numerically,  we take $0^+=0.02$ in Eq.(\ref{spectral}). }
\end{figure}

\subsection*{Two-particle spectral functions}
The two-particle spectral functions of a general superfluid state can be also defined as
\begin{align}\label{spectral1}
  &A_{\alpha\beta,\gamma\delta}(\textbf{q},\omega)=\sum_n\langle0|\Delta_{\alpha \beta \textbf{q}}|n\rangle\langle n|\Delta^{\dag}_{\gamma\delta \textbf{q}}|0\rangle \delta(\omega-\omega_{n0})-\sum_n\langle0|\Delta^{\dag}_{\gamma\delta \textbf{q}}|n\rangle\langle n|\Delta_{\alpha\beta \textbf{q}}|0\rangle \delta(\omega+\omega_{n0}),\notag\\
   & B_{\alpha\beta,\gamma\delta}(\textbf{q},\omega)=\sum_n\langle0|\Delta_{\alpha \beta \textbf{q}}|n\rangle\langle n|\Delta_{\gamma\delta, -\textbf{q}}|0\rangle \delta(\omega-\omega_{n0})-\sum_n\langle0|\Delta_{\gamma\delta, -\textbf{q}}|n\rangle\langle n|\Delta_{\alpha\beta \textbf{q}}|0\rangle \delta(\omega+\omega_{n0}),
\end{align}
where $\omega_{n0}=E_n-E_0$, $E_n$ and $|n\rangle$  are eigenvalues and eigenstates of Hamiltonian, respectively. $\Delta_{\alpha\beta \textbf{q}}$ is pairing fluctuation operator of superfluid order parameters, i.e.,
\begin{align}
   \Delta_{\alpha \beta \textbf{q}}=-U\sum_\textbf{k} \psi_{\alpha \textbf{k}+\textbf{q}}\psi_{\beta -\textbf{k}},
   \label{order}
\end{align}
where $\psi_{\alpha \textbf{k}+\textbf{q}}$ fermion field operator for single particle state $\alpha$ in momentum space, which satisfies  anti-commutation relation
\begin{align}
\psi_{\alpha \textbf{k}}\psi^{\dag}_{\beta \textbf{k}'}+\psi^{\dag}_{\beta \textbf{k}'}\psi_{\alpha \textbf{k}}=\delta_{\alpha\beta}\delta_{\textbf{k}\textbf{k}'}.
\end{align}

The two-particle normal (anomalous) Green functions can be expressed in terms of spectral functions, i.e.,
\begin{align}
  &G_{II\alpha\beta,\gamma\delta}(\textbf{q},i\omega_n)=\int_{-\infty}^{+\infty} d\omega\frac{A_{\alpha\beta,\gamma\delta}(\textbf{q},\omega)}{i\omega_n-\omega},\notag\\
   &F_{II\alpha\beta,\gamma\delta}(\textbf{q},i\omega_n)=\int_{-\infty}^{+\infty} d\omega\frac{B_{\alpha\beta,\gamma\delta}(\textbf{q},\omega)}{i\omega_n-\omega}.
\end{align}

Similarly as single-particle spectral function \cite{White1991}, the two-particle spectral functions satisfies a sum-rule, i.e.,
\begin{align}
 &\int d\omega A_{\alpha\beta,\gamma\delta}(\textbf{q},\omega)=\langle0|[\Delta_{\alpha\beta \textbf{q}},\Delta^{\dag}_{\gamma\delta \textbf{q}}] |0\rangle =U^2\langle0|\sum_\textbf{k}[\delta_{\beta\delta}\psi_{\alpha \textbf{k}}\psi^{\dag}_{\gamma \textbf{k}}-\delta_{\alpha\gamma}\psi^{\dag}_{\delta \textbf{k}}\psi_{ \beta \textbf{k}}]|0\rangle\notag\\
 &=U^2\langle0|\sum_\textbf{k}[\delta_{\beta\delta}\delta_{\alpha\gamma}-\delta_{\beta\delta}\psi^{\dag}_{\gamma \textbf{k}}\psi_{\alpha \textbf{k}}-\delta_{\alpha\gamma}\psi^{\dag}_{\delta \textbf{k}}\psi_{ \beta \textbf{k}}]|0\rangle,\notag\\
&\int d\omega B_{\alpha\beta,\gamma\delta}(\textbf{q},\omega)=
\langle0|[\Delta_{\alpha\beta \textbf{q}},\Delta_{\gamma\delta -\textbf{q}}] |0\rangle=0.
\end{align}

Taking the diagonal elements ($\gamma=\alpha;\delta=\beta$), 
\begin{align}
 &\int d\omega A_{\alpha\beta,\alpha\beta}(\textbf{q},\omega)=\langle0|[\Delta_{\alpha\beta \textbf{q}},\Delta^{\dag}_{\alpha\beta \textbf{q}}] |0\rangle =U^2\langle0|\sum_\textbf{k}[\psi_{\alpha \textbf{k}}\psi^{\dag}_{\alpha \textbf{k}}-\psi^{\dag}_{\beta \textbf{k}}\psi_{ \beta \textbf{k}}]|0\rangle\notag\\
&=U^2\langle0|\sum_\textbf{k}[1-\psi^{\dag}_{\alpha \textbf{k}}\psi_{\alpha \textbf{k}}-\psi^{\dag}_{\beta \textbf{k}}\psi_{ \beta \textbf{k}}]|0\rangle=U^2[N_{cell}-(N_\alpha+N_\beta)],
\end{align}
where $N_{cell}$ is the number of unit cells, $N_\alpha$ is the particle number of $\alpha-$th component.

In our dice lattice case, there exist three order parameters and they can be identified as $\Delta_{A(B /C)\downarrow A(B /C)\uparrow}=\Delta_{A(B / C)}$. The sum-rule is
\begin{align}
&\int d\omega B_{i,j}(\textbf{q},\omega)=
\langle0|[\Delta_{i \textbf{q}},\Delta_{j -\textbf{q}}] |0\rangle=0,\notag\\
&\int d\omega A_{i,j}(\textbf{q},\omega)=\langle0|[\Delta_{i \textbf{q}},\Delta^{\dag}_{j \textbf{q}}] |0\rangle=U^2[N_{cell}-(N_{i\downarrow}+N_{i\uparrow})]\delta_{ij}=U^2[N_{cell}-N_i]\delta_{ij},
\label{sumrule}
\end{align}
where indices $i,j=A,B,C$, particle number of $i-$th sublattice  $N_{i}=N_{i\uparrow}+N_{i\downarrow}$.

If we take the diagonal elements of $A_{ij}$ and sum over $i$, then we get a much more useful sum-rule
\begin{align}
&A\equiv\int d\omega \sum_{i}A_{i,i}(\textbf{q},\omega)=\sum_{i}\langle0|[\Delta_{i \textbf{q}},\Delta^{\dag}_{i \textbf{q}}] |0\rangle=U^2\sum_{i}[N_{cell}-N_i]=U^2[3N_{cell}-N]\propto -(n-3),
\label{sumrule1}
\end{align}
where total particle number $N=N_A+N_B+N_c$ and filling factor $n=N/N_{cell}$. It should be emphasized that the above sum-rule of spectral functions is exact, which does not depend on what state the system is. One can use the above sum-rule to check the calculations of various theories.

\begin{figure}
\begin{center}
\includegraphics[width=12.0cm]{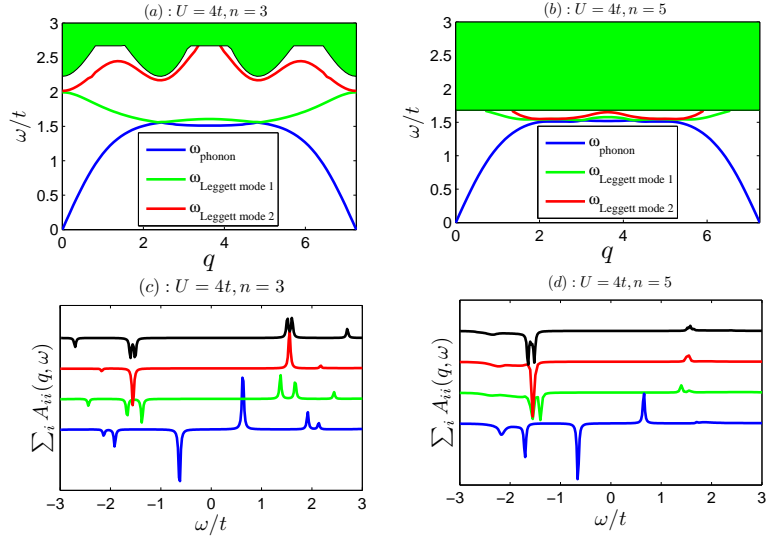}
\end{center}
\caption{ Collective modes [panel (a) and (b)] in x-direction $\textbf{q}=q \hat{x}$ and two-particle spectral functions [panel (c) and (d)]. (a): The phonon (Leggett) mode is plotted in blue (green and red) lines.  The interaction $U=4$, filling factor $n=3$ in panel (a).
(b): The phonon mode and Leggett modes with $U=4;n=5$.  (c): Spectral functions with $U=4;n=3$.  (d): Spectral functions with $U=4;n=5$. The spectral functions of $q=0.5, 1.5, 2.5$ and $3.5$ are plotted in blue, green, red and black lines, respectively.
}
\label{schematic3}
\end{figure}

The above Eq.(\ref{sumrule1}) shows that the integral of spectral functions is proportional to filling factor $n-3$ (relative to the half filling). If the filling factor is half filling ($n-3=0$), the integral of spectral functions is exactly zero. While when filling factor is larger (smaller) than half-filling [particle (hole)-doped case], the integral of spectral functions is negative (positive). Our numerical calculations verify the above observations. For example, when the filling is half-filling ($n=3$),  we see that the spectral functions are odd functions of frequency and the integral of spectral functions vanishes exactly [see panels (c)(d) of Fig. 4 and panel (c) of Fig. 5]. Panel (d) of Fig. 5 shows the spectral functions for particle-doped case ($n=5$), the weight of negative frequency $\omega<0$ is dominant, and the integral of spectral functions is negative.
The above findings may be useful to distinguish  between the hole-doped and particle-doped superfluid or superconductor in experiments.

In addition, for a fixed $\textbf{q}$, when $\omega$ approaches some non-degenerate collective modes (e.g., phonon or Leggett modes), from the definition Eq.(\ref{spectral1}), the spectral functions should take the following forms
\begin{align}
&A_{i,i}(\textbf{q},\omega)=\langle0|\Delta_{i \textbf{q}}|n\rangle\langle n|\Delta^{\dag}_{i\textbf{q}}|0\rangle \delta(\omega-\omega_{n0}),\notag\\
&A_{i,j}(\textbf{q},\omega)=\langle0|\Delta_{i \textbf{q}}|n\rangle\langle n|\Delta^{\dag}_{j\textbf{q}}|0\rangle \delta(\omega-\omega_{n0}),\notag\\
&A_{j,j}(\textbf{q},\omega)=\langle0|\Delta_{j \textbf{q}}|n\rangle\langle n|\Delta^{\dag}_{j\textbf{q}}|0\rangle \delta(\omega-\omega_{n0}),
\end{align}
where $|n\rangle=|\textbf{q}\rangle$ is wave function of the collective mode, $\omega_{n0}$ is its excitation energy.
 As a result of that, when $\omega\simeq \omega_{n0}$, the heights of peaks of two-particle spectral functions near the collective mode should satisfy a relation, i.e.,
\begin{align}
   |A_{i,j}(\textbf{q},\omega)|^2= A_{i,i}(\textbf{q},\omega)A_{j,j}(\textbf{q},\omega).\label{phonon1}
\end{align}

When $q$ is small ($q\rightarrow0$),  the matrix elements  of two-particle Green functions of zero frequency $\omega=0$ are proportional to product of two order parameters, namely \cite{Zhangyicai2020}
\begin{align}
  &G_{II,ij}(\textbf{q},0)\simeq  -4Z\Delta_{i}\Delta^*_{j},\notag\\
   & F_{II,ij}(\textbf{q},0)\simeq4Z\Delta_{i}\Delta_{j},
\end{align}
where $Z\equiv\sum_n[\frac{|\langle0|\hat{\theta}_{\textbf{q}}|n\rangle|^2}{\omega_{n0}}+\frac{|\langle0|\hat{\theta}_{-\textbf{q}}|n\rangle|^2}{\omega_{n0}}]\simeq1/(q^2\rho_s)>0$  is a  real number and $\hat{\theta}_{\textbf{q}}$ the phase operator \cite{zhangyicai2018,Lifshitz}.
\begin{table}
\begin{center}
\begin{tabular}{|c|c|c|c|c|c|c|c|c|c|c|}
\hline
1&2&3&4&5&6&7&8&9 \tabularnewline
\hline
 q&$\omega$& $A_{11}$ &  $A_{12}$ & $A_{22}$ &$\frac{A_{11}}{\Delta_{1}^{2}}$& $\frac{A_{12}}{\Delta_{1}\Delta_2}$& $\frac{A_{22}}{\Delta_{2}^{2}}$ & $\frac{A_{12}^2}{A_{11}A_{22}}$  \tabularnewline
 \hline
 0.01&0.018& 1274.5 &  1618.3 & 2054.8 &2306.5&2318.3 & 2330.0 & 1.00002  \tabularnewline
 \hline
 0.1& 0.15&129.4 & 167.9 & 217.8 &234.2&240.5 & 247.0 & 1.0002  \tabularnewline
 \hline
 0.5&0.66 &24.17 & 34.01 & 47.82 &43.75&48.72 & 54.22 & 1.0007  \tabularnewline
 \hline
\end{tabular}
\end{center}
\caption{The verifications of Eq.(\ref{phonon1}) and Eq.(\ref{phonon}) of spectral functions near the phonon modes with $U=4$ and filling factor $n=5$ [same as that in panel (b) and (d) of Fig. 5].
}
\end{table}
In such a case, it is expected that, for small $q$, the two-particle spectral functions near the phonon states [$\omega\simeq\omega_{phonon}(\textbf{q})$ or $\omega\simeq-\omega_{phonon}(-\textbf{q})$] are also proportional to product of superfluid order parameters, i.e.,
\begin{align}
  &A_{ij}(\textbf{q},\omega)\simeq 4Z_+ \Delta_{i}\Delta^*_{j}\delta[\omega-\omega_{phonon}(\textbf{q})]-4Z_- \Delta_{i}\Delta^*_{j}\delta[\omega+\omega_{phonon}(-\textbf{q})],\notag\\
    &B_{i,j}(\textbf{q},\omega)\simeq -4Z_+ \Delta_{i}\Delta_{j}\delta[\omega-\omega_{phonon}(\textbf{q})]+4Z_- \Delta_{i}\Delta_{j}\delta[\omega+\omega_{phonon}(-\textbf{q})],\label{phonon}
\end{align}
where  $Z_+=|\langle0|\hat{\theta}_{\textbf{q}}|\textbf{q}\rangle|^2$ and $Z_-=|\langle0|\hat{\theta}_{-\textbf{q}}|-\textbf{q}\rangle|^2$ are two real number, which are approximately equal as $q\rightarrow0$.
 
Our numerical calculations indeed verify  Eq.(\ref{phonon1})  and Eq.(\ref{phonon}) for the case of $q\rightarrow0$ . For example, some numerical results of spectral functions are listed in TABLE I.
It is found that, for extremely small wave vector ($q=0.01$), both Eqs.(\ref{phonon1}) and Eq.(\ref{phonon}) are satisfied approximately. As $q$ increases from $q=0.01$ to $q=0.1$ and eventually $q=0.5$,  Eq.(\ref{phonon}) becomes invalid gradually (see the columns $6,7,8$ in the TABLE I). Nevertheless, Eq.(\ref{phonon1}) still holds (see the last column of the TABLE I).

\section*{Discussion}

In conclusion, we investigate the superfluid properties of attractive Hubbard model in dice lattice. The order parameters and superfluid density as functions of filling factors and interactions are analyzed in great detail.
We emphasize the important roles of effective masses in the understanding of the asymptotic behaviors of superfluid density, especially for small filling factor. When both the filling factor and interaction are small, the superfluid density is given by the effective mass of single particle energy band.  When interaction is strong, the superfluid density is inversely proportional to interaction, which is also related to effective mass of tightly bound pairs. At strong interaction limit, the asymptotic behavior of superfluid density is captured by a parabola function of filling factor. In addition, the flat band has significant influences on superfluidity. To be specific, the infinite large density of states of flat band results in a linear interaction-dependence of superfluid order parameters. When interaction is weak, the existences of Dirac points and vanishingly small order parameters cause a logarithmic singularity of superfluid density.

 Due to the existence of three order parameters, there are three  collective modes, i.e., one is the gapless phonon, which corresponds to total density oscillations; the others are gapped Leggett modes, which are characterized by the relative density oscillations between different sublattices. It is found that the undamped Leggett modes can exist in strong interaction cases.
In addition, the collective modes can be reflected by sharp peaks in the two-particle spectral functions. The behaviors of two-particle spectral functions near the phonon modes are also analyzed in detail.
In addition, an exact sum-rule of spectral functions is  derived. For theoretical aspects, the sum-rule can be used to check the various theoretical calculations. In experiments, the sum-rule of the spectral functions may be useful to distinguish between the hole-doped and particle-doped superfluid (or superconductor).

\section*{Methods}
\subsection{Mean field theory}
The Hamiltonian Eq.(\ref{hamiltonian}) can be solved with the mean-field theory. The superfluid order parameter (pairing gap) $\Delta_{A(B,C)}=-U\langle \psi_{iA(B/C)\downarrow}\psi_{iA(B/C)\uparrow}\rangle$. The Bogoliubov-de Gennes Hamiltonian:
\begin{align}
\mathcal{H}_{\rm BdG}=\sum_{\bf k}\Psi^\dagger({\bf k})H_{\rm BdG}({\bf k})\Psi({\bf k}),
\end{align}
which $\Psi^\dagger({\bf k})\!\!=\!\![\psi_{A\uparrow}^\dagger({\bf k}),\psi_{B\uparrow}^\dagger({\bf k}),\psi_{C\uparrow}^\dagger({\bf k}),\psi_{A\downarrow}(-{\bf k}),\psi_{B\downarrow}(-{\bf k}),\psi_{C\downarrow}(-{\bf k})]$ and
\begin{align}
H_{\rm BdG}(\textbf{k})\!\!=\!\!\left[\!\!\begin{array}{cccccc} \begin{smallmatrix}
-\mu & h_{12}({\bf k}) & 0 &\Delta_{\rm A} & 0  & 0\\
h_{21}({\bf k}) & -\mu &h_{23}({\bf k}) & 0 & \Delta_{\rm B} & 0\\
0 & h_{32}({\bf k}) & -\mu & 0 & 0 & \Delta_{\rm C}\\
\Delta^*_{\rm A} & 0 & 0 &+\mu & -h_{12}^*(-{\bf k}) & 0\\
0 & \Delta^*_{\rm B} & 0 & -h_{21}^*(-{\bf k}) & \mu &-h^*_{23}(-{\bf k})\\
0 & 0 & \Delta^*_{\rm C} & 0 & -h_{32}^*(-{\bf k}) & +\mu
  \end{smallmatrix} \end{array}\!\!\right].
\end{align}
The Bogoliubov transformation of above equation is readily carried out to obtain the eigenenergies $E^{1,2,3+(-)}({\bf k})$, where $E^{n-}(\textbf{k})=-E^{n+}(-\textbf{k})<0$ and $1,2,3$ represent three branches of energy.
The thermodynamic potential per unit cell in the ground state is
\begin{align}
    \begin{split}
    	\Omega=&\frac{1}{N_{cell}}\sum_{\bf k}[E^{1-}({\bf k})+E^{2-}({\bf k})+E^{3-}({\bf k})-3\mu]+\frac{(\Delta_{\rm A}^2+\Delta_{\rm B}^2+\Delta_{\rm C}^2)}{U},
    \end{split}
    \end{align}
where $N_{cell}$ is the number of unit cells.
The order parameters (or pairing gaps) and chemical potential  can be obtained by solving $\partial \Omega/\partial \Delta_{\rm A,B,C}=0$, and $n=-\partial\Omega/\partial\mu$ consistently. In the whole paper, we set $t=1$ as the energy unit and lattice spacing $a=1$.
\subsection{Asymptotic behaviors of superfluid density }
For dice lattices,  the superfluid order parameters for three sublattices are
 \begin{align}
&\Delta_{A}=-U\langle \psi_{iA\downarrow}\psi_{iA\uparrow}\rangle,\notag\\
&\Delta_{B}=-U\langle \psi_{iB\downarrow}\psi_{iB\uparrow}\rangle,\notag\\
&\Delta_{C}= -U\langle \psi_{iC\downarrow}\psi_{iC\uparrow}\rangle,
\end{align}

 Using the mean-field decoupling, the  BdG Hamiltonian
is
\begin{eqnarray}
\mathcal{H}_{\rm BdG}=\sum_{\bf k}\Psi^\dagger({\bf k})H_{\rm BdG}({\bf k})\Psi({\bf k}),
\end{eqnarray}
where $\Psi^\dagger({\bf k})\!\!=\!\![\psi_{A\uparrow}^\dagger({\bf k}),\psi_{B\uparrow}^\dagger({\bf k}),\psi_{C\uparrow}^\dagger({\bf k}),\psi_{A\downarrow}(-{\bf k}),\psi_{B\downarrow}(-{\bf k}),\psi_{C\downarrow}(-{\bf k})]$  and
\begin{eqnarray}
H_{\rm BdG}({\bf k})=\left( \!\!\!                
 \begin{array}{cccccc}   
  h^{p}_{11}(\textbf{k})& h^{p}_{12}(\textbf{k}) & 0& \Delta_A   &0     & 0 \\  
  h^{p}_{21}(\textbf{k})  & h^{p}_{22}(\textbf{k})  &h^{p}_{23}(\textbf{k}) & 0&\Delta_B  & 0  \\  
   0  & h^{p}_{23}(\textbf{k})  &h^{p}_{33}(\textbf{k}) & 0  & 0  &\Delta_C\\  
   \Delta^{*}_{A} &0 & 0 &-h^{h}_{11}(-\textbf{k}) & -h^{h}_{12}(-\textbf{k})&0 \\  
  0 &\Delta^{*}_{B} & 0 &-h^{h}_{21}(-\textbf{k}) & -h^{h}_{22}(-\textbf{k})&-h^{h}_{23}(-\textbf{k}) \\  
   0 &0 & \Delta^{*}_{C} &0 & -h^{h}_{32}(-\textbf{k})&-h^{h}_{33}(-\textbf{k}) \\  
\end{array}\!\!\!\right),
\end{eqnarray}
where the particle and hole  part Hamiltonian are
\begin{eqnarray}
h^{p}(\textbf{k})=\left( \!\!\!                
 \begin{array}{ccc}   
h^{p}_{11}(\textbf{k})& h^{p}_{12}(\textbf{k}) & 0  \\  
h^{p}_{21}(\textbf{k})  & h^{p}_{22}(\textbf{k})  &h^{p}_{23}(\textbf{k})\\
  0  & h^{p}_{23}(\textbf{k})  &h^{p}_{33}(\textbf{k})   \\  
\end{array}\!\!\!\right),
\end{eqnarray}
and
\begin{eqnarray}
h^{h}(-\textbf{k})=\left( \!\!\!                
 \begin{array}{cccc}   
 h^{h}_{11}(-\textbf{k}) & h^{h}_{12}(-\textbf{k})&0  \\  
 h^{h}_{21}(-\textbf{k}) & h^{h}_{22}(-\textbf{k})&h^{h}_{23}(-\textbf{k})  \\  
 0 & h^{h}_{32}(-\textbf{k})&h^{h}_{33}(-\textbf{k})\\
\end{array}\!\!\!\right),
\end{eqnarray}
respectively.
The above BdG matrix can be diagonalized, e.g.,
\begin{eqnarray}
&H_{\rm BdG}({\bf k})|n+,k\rangle=E^{n+}(\textbf{k})|n+,\textbf{k}\rangle, \notag\\
&H_{\rm BdG}({\bf k})|n-,k\rangle=E^{n-}(\textbf{k})|n-,\textbf{k}\rangle,
\end{eqnarray}
where quasi-particle energy  $E^{n+}(\textbf{k})=E^{n}(\textbf{k}))>0$ and $E^{n-}(\textbf{k})=-E^{n+}(-\textbf{k})<0$ due to particle-hole symmetry of the Bogoliubov-de Gennes Hamiltonian. For three band dice lattices, the indices $n=1,2,3$.
Eigen-states $|n+,\textbf{k}\rangle$ and $|n-,\textbf{k}\rangle$ (``quasi-particle basis")  can be also written as
\begin{eqnarray}
&|n+,\textbf{k}\rangle=\sum_{s=1,2,3}W^{+}_{ns}(\textbf{k})|ps,\textbf{k}\rangle +\sum_{t=1,2,3}W^{+}_{nt}(\textbf{k})|ht,\textbf{k}\rangle,\notag\\
&|n-,\textbf{k}\rangle=\sum_{s=1,2,3}W^{-}_{ns}(\textbf{k})|ps,\textbf{k}\rangle +\sum_{t=1,2,3}W^{-}_{nt}(\textbf{k})|ht,\textbf{k}\rangle,
\end{eqnarray}
where $|ps,\textbf{k}\rangle $ and $|ht,\textbf{k}\rangle $ are eigen-states of particle and hole part Hamiltonian, respectively, i.e,
\begin{eqnarray}
&h^{p}(\textbf{k})|ps,\textbf{k}\rangle=\epsilon^{p}_s(\textbf{k})|ps,\textbf{k}\rangle,\notag\\
&h^{h}(-\textbf{k})|ht,\textbf{k}\rangle=\epsilon^{h}_t(\textbf{k})|ht,\textbf{k}\rangle.
\end{eqnarray}
They  are also the ``energy band basis" for single-particle Hamiltonian. Then $W^{+}$ and $W^{-}$ are transformation matrices between the above two sets of basis. Due to the time reversal symmetry in dice lattices, $h^{p}(\textbf{k})=h^{h}(-\textbf{k})= h(\bf{k})$ [ Eq.(\ref{H0})], $\epsilon^{p}_n(\textbf{k})=\epsilon^{h}_n(\textbf{k})\equiv \epsilon_n(\textbf{k})$ and $|pn,\textbf{k}\rangle=|hn,\textbf{k}\rangle\equiv |n,\textbf{k}\rangle$.

The thermodynamic potential per unit cell is
\begin{align}
    \begin{split}
    	\Omega=&\frac{1}{N_{cell}}\sum_{\bf k}[E^{1-}(\textbf{k})+E^{2-}(\textbf{k})+E^{3-}(\textbf{k})]-3\mu+\frac{(\Delta_{\rm A}^2+\Delta_{\rm B}^2+\Delta_{\rm C}^2)}{U}=\frac{1}{N_{cell}}\sum_{\textbf{k},n=1,2,3}E^{n-}(\textbf{k})+Cons,
    \end{split}
    \end{align}
where $N_{cell}$ is the number of unit cells. The constant $Cons=-3\mu+\frac{(\Delta_{\rm A}^2+\Delta_{\rm B}^2+\Delta_{\rm C}^2)}{U}$.

The superfluid density can be obtained with  phase twist method. Assuming the superfluid order parameters undergo a phase variation, e.g, $\triangle(\textbf{r})\rightarrow \Delta(\textbf{r})e^{2i\textbf{q}\cdot\textbf{r}}$,
the superfluid density  tensor $\rho_{sij}$ can be written as
 \begin{eqnarray}
 \rho_{sij}= \frac{\partial^2 \Omega(\textbf{q})}{\partial q_i \partial q_i}|_{q\rightarrow0},
\end{eqnarray}
where $\Omega(\textbf{q})$ is thermodynamical potential.
In the presence of phase twist, the BdG Hamiltonian (after applying a gauge transformation) becomes
\begin{eqnarray}
 H_{\rm BdG}= \sum_{\bf{k}} \Psi^{\dag}(\textbf{k}) H_{\rm BdG}({\bf k,q})\Psi(\textbf{k}),
\end{eqnarray}
where $\Psi^\dagger({\bf k})\!\!=\!\![\psi_{A\uparrow}^\dagger({\bf k}),\psi_{B\uparrow}^\dagger({\bf k}),\psi_{C\uparrow}^\dagger({\bf k}),\psi_{A\downarrow}(-{\bf k}),\psi_{B\downarrow}(-{\bf k}),\psi_{C\downarrow}(-{\bf k})]$  and the BdG matrix
\begin{eqnarray}
&&H_{\rm BdG}({\bf k,q})=\left( \!\!\!                
 \begin{array}{cccccc}   
  h^{p}_{11}(\textbf{k}+\textbf{q})& h^{p}_{12}(\textbf{k}+\textbf{q}) & 0& \Delta_A   &0     & 0 \\  
  h^{p}_{21}(\textbf{k}+\textbf{q})  & h^{p}_{22}(\textbf{k}+\textbf{q})  &h^{p}_{23}(\textbf{k}+\textbf{q}) & 0&\Delta_B  & 0  \\  
   0  & h^{p}_{23}(\textbf{k}+\textbf{q})  &h^{p}_{33}(\textbf{k}+\textbf{q}) & 0  & 0  &\Delta_C\\  
   \Delta^{*}_{A} &0 & 0 &-h^{h}_{11}(-\textbf{k}+\textbf{q}) & -h^{h}_{12}(-\textbf{k}+\textbf{q})&0 \\  
  0 &\Delta^{*}_{B} & 0 &-h^{h}_{21}(-\textbf{k}+\textbf{q}) & -h^{h}_{22}(-\textbf{k}+\textbf{q})&-h^{h}_{23}(-\textbf{k}+\textbf{q}) \\  
   0 &0 & \Delta^{*}_{C} &0 & -h^{h}_{32}(-\textbf{k}+\textbf{q})&-h^{h}_{33}(-\textbf{k}+\textbf{q}) \\  
\end{array}\!\!\!\right).
\end{eqnarray}
After diagonalizing the above BdG matrix, we can get the eigen-energies $E^{n+}(\bf{k,q})$ and $E^{n-}(\bf{k,q})$.
The thermodynamical potential becomes
\begin{align}
    	\Omega(\textbf{q})=\frac{1}{N_{cell}}\sum_{\textbf{k},n=1,2,3}  E^{n-}(\textbf{k},\textbf{q})+Cons.
    \end{align}
    
In addition, if a Hamiltonian $H$, its eigen-states $|n\rangle$ and eigen-energies $E^n$ depend on parameters $\textbf{k}$ and $\textbf{q}$,  they should satisfy
\begin{eqnarray}
 &H|n\rangle=E^{n}|n\rangle,\notag\\
 & \langle m|n\rangle=\delta_{mn},\notag\\
 &E^n=\langle n |H|n\rangle,\notag\\
 &\frac{\partial E^n}{\partial q_i}=\langle n|\frac{\partial H}{\partial q_i}|n\rangle,\notag\\
 &\frac{\partial E^n}{\partial k_i}=\langle n|\frac{\partial H}{\partial k_i}|n\rangle,\notag\\
 &\langle m|\frac{\partial n}{\partial q_i}\rangle=\frac{\langle m|\frac{\partial H}{\partial q_i}|n\rangle}{E^n-E^m},\notag\\
 &\langle m|\frac{\partial n}{\partial k_i}\rangle=\frac{\langle m|\frac{\partial H}{\partial k_i}|n\rangle}{E^n-E^m},\notag\\
  &\frac{\partial^2 E^n}{\partial q_i \partial q_j}=\langle n|\frac{\partial^2 H}{\partial q_i \partial q_j}|n\rangle+\sum_{m\neq n}[\frac{\langle n|\frac{\partial H}{\partial q_i}|m\rangle\langle m|\frac{\partial H}{\partial q_j}|n\rangle}{E^n-E^m}+\frac{\langle n|\frac{\partial H}{\partial q_j}|m\rangle\langle m|\frac{\partial H}{\partial q_i}|n\rangle}{E^n-E^m}],\notag\\
 &\frac{\partial^2 E^n}{\partial k_i \partial k_j}=\langle n|\frac{\partial^2 H}{\partial k_i \partial k_j}|n\rangle+\sum_{m\neq n}[\frac{\langle n|\frac{\partial H}{\partial k_i}|m\rangle\langle m|\frac{\partial H}{\partial k_j}|n\rangle}{E^n-E^m}+\frac{\langle n|\frac{\partial H}{\partial k_j}|m\rangle\langle m|\frac{\partial H}{\partial k_i}|n\rangle}{E^n-E^m}].
\end{eqnarray}
Due to $\langle n|\frac{\partial^2 H_{\rm BdG}({\bf k,q})}{\partial k_i \partial k_j}|n\rangle=\langle n|\frac{\partial^2 H_{\rm BdG}({\bf k,q})}{\partial q_i \partial q_j}|n\rangle$ for the BdG matrix $H_{\rm BdG}({\bf k,q})$, so the superfluid density can be written as
\begin{eqnarray}
 &&\rho_{sij}=\frac{1}{N_{cell}}\sum_{\textbf{k}nm}\frac{\partial^2 E^{n-}}{\partial k_i \partial k_j}|_{q\rightarrow 0}\notag\\
 &&+\frac{1}{N_{cell}}\sum_{\textbf{k}nm}\{[\frac{\langle n-|\frac{\partial H}{\partial q_i}|m+\rangle\langle m+|\frac{\partial H}{\partial q_j}|n-\rangle}{E^{n-}-E^{m+}}+\frac{\langle n-|\frac{\partial H}{\partial q_j}|m+\rangle\langle m+|\frac{\partial H}{\partial q_i}|n-\rangle}{E^{n-}-E^{m+}}]\notag\\
 &&-[\frac{\langle n-|\frac{\partial H}{\partial k_i}|m+\rangle\langle m+|\frac{\partial H}{\partial k_j}|n-\rangle}{E^{n-}-E^{m+}}+\frac{\langle n-|\frac{\partial H}{\partial k_j}|m+\rangle\langle m+|\frac{\partial H}{\partial k_i}|n-\rangle}{E^{n-}-E^{m+}}]\}_{q\rightarrow0},
\end{eqnarray}
where we use $H$ to denote the BdG matrix $H_{\rm {BdG}}(\bf{k,q})$. $|m+\rangle\equiv |m+,\textbf{k},\textbf{q}\rangle$ ($|n-\rangle\equiv |n-,\textbf{k},\textbf{q}\rangle$) and $E^{m+}\equiv E^{m+}(\textbf{k},\textbf{q})[E^{n-}\equiv E^{n-}(\textbf{k},\textbf{q})]$ are eigen-states and eigen-energies of  Hamiltonian $H_{\rm {BdG}}(\bf{k,q})$. 
The first term of superfluid depends on the second derivative $\frac{\partial^2 E^{n-}}{\partial k_i \partial k_j}$,  it would vanish. Because  quasi-particle energy $E^{n-}(k_x,k_y)$ is a periodic function of $k_x$, the summation over $k$ can be transformed into an integral. An  integral of a  derivative of periodic function over a period is zero.
So the superfluid density is
\begin{eqnarray}
 &\rho_{sij}= \frac{\partial^2 \Omega(\textbf{q})}{\partial q_i \partial q_i}=\frac{1}{N_{cell}}\sum_{\textbf{k}nm}[\frac{\langle n-|\frac{\partial H}{\partial q_i}|m+\rangle\langle m+|\frac{\partial H}{\partial q_j}|n-\rangle}{E^{n-}-E^{m+}}-\frac{\langle n-|\frac{\partial H}{\partial k_i}|m+\rangle\langle m+|\frac{\partial H}{\partial k_j}|n-\rangle}{E^{n-}-E^{m+}}+(i\leftrightarrow j)]_{q\rightarrow0}.
\end{eqnarray} 

When $q\rightarrow0$,
\begin{align}
\frac{\partial H}{\partial q_i}=\left( \!\!\!                
 \begin{array}{cccc}   
[\frac{\partial h^{p}(\textbf{k})}{\partial k_i}]_{3\times3} &(0)_{3\times3}  \\  
 (0)_{3\times3}  & [\frac{\partial h^{h}(-\textbf{k})}{\partial k_i}]_{3\times3}  \\  
\end{array}\!\!\!\right),    
    \end{align}	
and \begin{eqnarray}
\frac{\partial H}{\partial k_i}=\left( \!\!\!                
 \begin{array}{cccc}   
[\frac{\partial h^{p}(\textbf{k})}{\partial k_i}]_{3\times3} &(0)_{3\times3}  \\  
 (0)_{3\times3}  & -[\frac{\partial h^{h}(-\textbf{k})}{\partial k_i}]_{3\times3}  \\  
\end{array}\!\!\!\right).
\end{eqnarray}
Using the energy-band basis, the superfluid density           \begin{eqnarray}
 &&\rho_{sij}=\frac{2}{N_{cell}}\sum_{\textbf{k}nm s_1s_2t_1t_2}[\frac{W^{-*}_{ns_1}W^{+}_{ms_2}W^{+*}_{mt_1}W^{-}_{nt_2}}{E^{n-}-E^{m+}}\langle s_1p|\frac{\partial h^{p}}{\partial k_i}|s_2 p\rangle\langle t_1 h|\frac{\partial h^{h}}{\partial k_j}|t_2 h\rangle\notag\\
 &&+\frac{W^{-*}_{nt_1}W^{+}_{mt_2}W^{+*}_{ms_1}W^{-}_{ns_2}}{E^{n-}-E^{m+}}\langle t_1 h|\frac{\partial h^{h}}{\partial k_i}|t_2h\rangle\langle s_1p|\frac{\partial h^{p}}{\partial k_j}|s_2 p\rangle+(i\leftrightarrow j)].
\end{eqnarray}

When interaction $U\rightarrow\infty$, the chemical potential $\mu \propto U$ and the order parameters $\Delta_A\simeq\Delta_B\simeq\Delta_C\propto U$, which corresponds to the uniform pairing case \cite{Liang2017}. In such a case, the superfluidity takes place independently in the respective energy bands (no inter-band pairings).
 The transformation matrix takes the following form
\begin{eqnarray}
 &W^{+}_{ns}(\textbf{k})\simeq\delta_{ns}\sqrt{\frac{1}{2}(1+\frac{\epsilon_{n}(\textbf{k})}{E^{n}(\textbf{k})})},\notag\\
 &W^{+}_{nt}(\textbf{k})\simeq\delta_{nt}\sqrt{\frac{1}{2}(1-\frac{\epsilon_{n}(\textbf{k})}{E^{n}(\textbf{k})})}.
\end{eqnarray}
In addition, the particle-hole symmetry of BdG matrix implies that the transformation matrices satisfy 
\begin{eqnarray}
&W^{-}_{ns}(\textbf{k})=\delta_{st}\delta_{ns}W^{+*}_{nt}(\textbf{k})=\delta_{st}\delta_{ns}\sqrt{\frac{1}{2}(1-\frac{\epsilon_{n}(\textbf{k})}{E^{n}(\textbf{k})})},\notag\\
&W^{-}_{nt}(\textbf{k})=-\delta_{st}\delta_{ns}W^{+*}_{ns}(\textbf{k})=-\delta_{st}\delta_{ns}\sqrt{\frac{1}{2}(1+\frac{\epsilon_{n}(\textbf{k})}{E^{n}(\textbf{k})})}.
\end{eqnarray}
When $U\rightarrow\infty$, the order parameter and chemical potential
\begin{eqnarray}
& \Delta \simeq \frac{U}{6}\sqrt{n(6-n)},\notag\\
 &\mu \simeq \frac{U}{6}(n-3),
 \end{eqnarray}
 where $n$ is filling factor (particle number per unit cell).
In addition, the quasi-particle energies satisfy
\begin{align}
E^{n+}(\textbf{k})=-E^{n-}(-\textbf{k})=\sqrt{(\epsilon_{n}(\textbf{k}))^2+\Delta^2}\simeq\sqrt{\mu^2+\Delta^2};\quad E^{n-}-E^{m+}\simeq  -2\sqrt{\mu^2+\Delta^2},
\end{align}
where $\epsilon_{n}(\textbf{k})\simeq-\mu$ as $U\rightarrow \infty$. The superfluid density is reduced to
\begin{eqnarray}
& \rho_{sij}\simeq \frac{\Delta^2}{2N_{cell}(\mu^2+\Delta^2)^{3/2}}\sum_{\textbf{k}nm}[\langle n|\frac{\partial h(\bf{k})}{\partial k_i}|m\rangle\langle m|\frac{\partial h(\bf{k})}{\partial k_j}|n\rangle+\langle n|\frac{\partial h(\bf{k})}{\partial k_j}|m\rangle\langle m|\frac{\partial h(\bf{k})}{\partial k_i}|n\rangle],
\end{eqnarray}
where $|n\rangle \equiv |n,\textbf{k}\rangle$ is energy band basis, and $h(\bf{k})$ is single-particle Hamiltonian [ Eq.(\ref{H0})].

 So the superfluid density
\begin{eqnarray}
&& \rho_{sij}\simeq \frac{\Delta^2}{2N_{cell}(\mu^2+\Delta^2)^{3/2}}\sum_{\textbf{k}nm}[\langle n|\frac{\partial h(\bf{k})}{\partial k_i}|m\rangle\langle m|\frac{\partial h(\bf{k})}{\partial k_j}|n\rangle+\langle n|\frac{\partial h(\bf{k})}{\partial k_j}|m\rangle\langle m|\frac{\partial h(\bf{k})}{\partial k_i}|n\rangle],\notag\\
&&=\frac{2n(6-n)}{9U}S_{ij},
\end{eqnarray}
where\begin{eqnarray}
&& S_{ij}=\frac{1}{2N_{cell}}\sum_{\textbf{k}nm}[\langle n|\frac{\partial h(\bf{k})}{\partial k_i}|m\rangle\langle m|\frac{\partial h(\bf{k})}{\partial k_j}|n\rangle+\langle n|\frac{\partial h(\bf{k})}{\partial k_j}|m\rangle\langle m|\frac{\partial h(\bf{k})}{\partial k_i}|n\rangle],
\end{eqnarray}
is a rank-two tensor, which only depends on single-particle Hamiltonian. For dice lattices in our work, $S_{ij}=6t^2\delta_{ij}$.
So the asymptotic formula for superfluid density (particle number per unit cell) is obtained ($U\rightarrow\infty$)
  \begin{eqnarray}
 \rho_{sij}\simeq \frac{4t^2n(6-n)}{3U}\delta_{ij}.
\end{eqnarray}

\subsection{Gaussian fluctuation method}
To investigate the collective modes, it is very convenient to formulate the theory with functional integration. The partition function of grand canonical ensemble is written as functional integral of Grassman fields $\bar{\psi}$ and $\psi$
\begin{align}
Z \equiv \int D[\bar{\psi}_{iA\sigma}\bar{\psi}_{iB\sigma}\bar{\psi}_{iC\sigma}]D[\psi_{iA\sigma}\psi_{iB\sigma}\psi_{iC\sigma}] e^{-S},
\end{align}
where we have the action $S$ given by
\begin{align}\nonumber
S=\sum_{iA,\sigma;iB,\sigma;iC,\sigma}\int_{0}^{\beta}d\tau [\bar{\psi}_{iA\sigma}(\tau)\frac{\partial\psi_{iA\sigma}(\tau)}{\partial \tau}+\bar{\psi}_{iB\sigma}(\tau)\frac{\partial\psi_{iB\sigma}(\tau)}{\partial \tau}+\bar{\psi}_{iC\sigma}(\tau)\frac{\partial\psi_{iC\sigma}(\tau)}{\partial \tau}
+H(\psi_{iA\sigma},\psi_{iB\sigma},\psi_{iC\sigma})],
\end{align}
where the inverse temperature $\beta=1/T$. 
In this paper, we will only focus on the properties at zero temperature and will take $\beta\rightarrow\infty$ at the end.
By introducing the Hubbard-Stratanovich fields  $\Delta_{iA(B,C)}$ for sub-lattice \textbf{A(B,C)} as usual, we have
\begin{align}
Z =\int D\Delta^{*}D \Delta D \bar{\psi} D \psi e^{-S_\Delta},
 \end{align}
 and the effective action is given by
 \begin{align}
S_{\Delta } = \int_{0}^{\beta}d\tau \sum_{Ai,\sigma;Bj,\sigma}[\frac{|\Delta_{iA\sigma}(\tau)|^2}{U}+\frac{|\Delta_{iB\sigma}(\tau)|^2}{U}+\frac{|\Delta_{iC\sigma}(\tau)|^2}{U}- \bar{\psi} G^{-1}(\Delta_{iA}(\tau),\Delta_{iB}(\tau),\Delta_{iC}(\tau))\psi].
\end{align}

In the following, we will use the subscribe $i_1$, and $i'_{1}$ [($i_{2}$, $i'_{2}$)/($i_{3}$, $i'_{3}$)] to denote the  positions of sub-lattices \textbf{A} [\textbf{B}/\textbf{C}] unless stated otherwise.
With the basis $\bar{\psi}=[\bar{\psi}_{i_1A\uparrow},\bar{\psi}_{i_2B\uparrow},\bar{\psi}_{i_3C\uparrow};\psi_{i_1A\downarrow},\psi_{i_2B\downarrow},\psi_{i_3C\downarrow}]$, the inverse Nambu-Gorkov Green function $G^{-1}$ takes following form
\begin{align}
&G^{-1}=\left(                 
 \begin{array}{cccccc}   
   (A)_{i_1,i'_1}  &   (A)_{i_1,i_2}  &0   &   (B)_{i_1,i'_1}   &0  &  0\\  
  (A)^{*}_{i_1,i_2}    &   (A)_{i_2,i'_2} & (A)_{i_2,i_3}    &     0   &  (B)_{i_2,i'_2}&0\\  
  0    &   (A)^{*}_{i_2,i_3} & (A)_{i_3,i'_3}    &     0   &  0&(B)_{i_3,i'_3}\\  
   (B)^*_{i_1,i'_1}    &    0 &0   &    (D)_{i_1,i'_1}   &  (D)_{i_1,i_2}& 0\\  
   0    &     (B)^*_{i_2,i'_2} &0    &   (D)^{*}_{i_1,i_2}   &   (D)_{i_2,i'_2} &(D)_{i_2,i_3} \\
    0    &     0& (B)^*_{i_3,i'_3}    &   0   &   (D)^{*}_{i_2,i_3} &(D)_{i_3,i'_3} \\
\end{array}\right),
\end{align}
where \begin{align}
A_{i_1,i'_1} =&(-\partial_\tau+\mu)\delta_{i_1,i'_1}\delta(\tau-\tau'),\notag\\
A_{i_1,i_2}= &-t\sum_\delta\delta_{i_1,i_2-\textbf{a}_\delta}\delta(\tau-\tau'),\notag\\
A_{i_2,i'_2}= &(-\partial_\tau+\mu)\delta_{i_2,i'_2}\delta(\tau-\tau'),\notag\\
A_{i_2,i_3}= &-t\sum_\delta\delta_{i_2,i_3-\textbf{a}_\delta}\delta(\tau-\tau'),\notag\\
A_{i_3,i'_3}= &(-\partial_\tau+\mu)\delta_{i_3,i'_3}\delta(\tau-\tau'),\notag\\
B_{i_1,i'_1}=&\Delta_{i_1A}(\tau)\delta_{i_1,i'_1}\delta(\tau-\tau'),\notag\\
B_{i_2,i'_2}=&\Delta_{i_2B}(\tau)\delta_{i_2,i'_2}\delta(\tau-\tau'),\notag\\
B_{i_3,i'_3}=&\Delta_{i_3C}(\tau)\delta_{i_3,i'_3}\delta(\tau-\tau'),\notag\\
D_{i_1,i'_1}=&(-\partial_\tau-\mu)\delta_{i_1,i'_1}\delta(\tau-\tau'),\notag\\
D_{i_1,i_2}= &t\sum_\delta\delta_{i_1,i_2-\textbf{a}_\delta}\delta(\tau-\tau'),\notag\\
D_{i_2,i'_2}=&(-\partial_\tau-\mu)\delta_{i_2,i'_2}\delta(\tau-\tau'),\notag\\
D_{i_2,i_3}= &t\sum_\delta\delta_{i_2,i_3-\textbf{a}_\delta}\delta(\tau-\tau'),\notag\\
D_{i_3,i'_3}=&(-\partial_\tau-\mu)\delta_{i_3,i'_3}\delta(\tau-\tau').
\end{align}
 The functional integral is quadratic with respect to $\psi$ fields, we can integrate it out
\begin{align}
Z &= \int D\Delta^{*}D \Delta \exp \{-\int^{\beta}_{0} d\tau \sum_{i} [\frac{|\Delta_{iA}(\tau)|^2}{U}+\frac{|\Delta_{iB}(\tau)|^2}{U}+\frac{|\Delta_{iC}(\tau)|^2}{U}-{\rm Tr}\ln G^{-1}(\Delta_{iA}(\tau),\Delta_{iB}(\tau),,\Delta_{iC}(\tau))]\}.
\end{align}

Further   assuming $\Delta_{iA}$, $\Delta_{iB}$ and $\Delta_{iC}$ can be written as $\Delta_{iA}(\tau)=\Delta_A+\delta \Delta_{iA}(\tau)$, $\Delta_{iB}(\tau)=\Delta_B+\delta \Delta_{iB}(\tau)$ and $\Delta_{iC}(\tau)=\Delta_C+\delta \Delta_{iC}(\tau)$, where $\Delta_{A(B/C)}$ are not dependent on spatial and time variables.
It is convenient to write the $G^{-1}$ in momentum and Matsubara frequency spaces, i.e.,
\begin{align}
&G^{-1}(\textbf{k},\textbf{k}')=G^{-1}_{0}(\textbf{k},\textbf{k}')+K(\textbf{k},\textbf{k}'),
\end{align}
where
\begin{align}
&G_{0}^{-1}(\textbf{k},\textbf{k}')=\left(                 
 \begin{array}{cccccc}   
   A_{1,1}  &   A_{1,2}   &0 &   \Delta_A    & 0& 0\\  
  A^{*}_{1,2}    &   A_{2,2} & A_{2,3}    &     0 & \Delta_B  & 0 \\  
    0    &   A^{*}_{2,3} & A_{2,3}    &     0 & 0  & \Delta_C \\  
   \Delta^{*}_{A}     &  0&0    &    D_{1,1}   &  D_{1,2}&0\\  
   0    &     \Delta^{*}_{B}   &0  &   D^{*}_{1,2}   &  D_{2,2}&D_{2,3} \\
   0    &     0   &\Delta^{*}_{C}  &  0  &  D^{*}_{2,3}&D_{3,3} \\
\end{array}\right)\delta(\textbf{k}-\textbf{k}'),
\end{align}
and
\begin{align}
&K (\textbf{k},\textbf{k}')=\left( \!\!\!                
 \begin{array}{cccccc}   
   0  &   0    & 0&    \Delta_A(-k'+k)    & 0 &0\\  
  0    &   0     & 0&    0   &   \Delta_B(-k'+k)&0 \\  
  0    &   0     & 0&    0   &   0&\Delta_C(-k'+k) \\  
   \Delta_{A}^{*}(k'-k) &0&0    &    0    &    0   &  0\\  
   0    &     \Delta_{B}^{*}(k'-k)  &0   &   0  &   0&0 \\
    0    &     0  &\Delta_{C}^{*}(k'-k)   &   0  &   0&0 \\
\end{array} \!\!\!\right),
\end{align}
where we introduce $\delta(\textbf{k}-\textbf{k}')\equiv \delta^{2}(\textbf{k}-\textbf{k}')\delta_{n,n'}$, $k=(\textbf{k},i\omega_n)$, $k'=(\textbf{k}',i\omega_{n'})$, $\delta k=k'-k=(\textbf{k}'-\textbf{k},i\omega_{n'}-i\omega_n)$ and
\begin{align}
&A_{1,1}=i\omega_n+\mu,\notag\\
&A_{1,2}=-t\sum_\delta e^{i\textbf{k}\cdot \textbf{a}_\delta},\notag\\
&A_{2,2}=i\omega_n+\mu,\notag\\
&A_{2,3}=-t\sum_\delta e^{i\textbf{k}\cdot \textbf{a}_\delta},\notag\\
&A_{3,3}=i\omega_n+\mu,\notag\\
&D_{1,1}=i\omega_n-\mu,\notag\\
&D_{1,2}=t\sum_\delta e^{i\textbf{k}\cdot \textbf{a}_\delta},\notag\\
&D_{2,2}=i\omega_n-\mu,\notag\\
&D_{2,3}=t\sum_\delta e^{i\textbf{k}\cdot \textbf{a}_\delta},\notag\\
&D_{3,3}=i\omega_n-\mu.
\end{align}

 Expanding action $S_\Delta$ to second order of $\delta\Delta$ ,
one  gets the thermodynamic potential
\begin{align}
Z\approx e^{-S_0}\int D \bar{\eta}_q D\eta_q e^{-\delta S},
\end{align}
where  $S_0$ is the action from the mean-field contribution and $q=\delta k$.  The linear terms vanish due to the mean-field (saddle-point) equations. The fluctuation contribution is
\begin{align}
\delta S=\frac{1}{2}\sum_{\textbf{q},n}\bar{\eta_q}M\eta_q=\sum_{\textbf{q},n>0}\bar{\eta_q}M\eta_q,
\end{align}
where
\begin{align}
\bar{\eta}_\textbf{\emph{q}}=[\Delta^{*}_{A}(\textbf{q},i\omega_n),\Delta^{*}_{B}(\textbf{q},i\omega_n),\Delta^{*}_{C}(\textbf{q},i\omega_n);\Delta_{A}(-\textbf{q},-i\omega_n),\Delta_{B}(-\textbf{q},-i\omega_n),\Delta_{C}(-\textbf{q},-i\omega_n)]\notag
\end{align}
and the fluctuation matrix $M$ is given by
\begin{eqnarray}
 &&M_{ij}(\textbf{q},i\omega_n)=\frac{1}{\beta}\sum_{\textbf{k},n'}G^{0}_{ij}(k+q)G^{0}_{j+m,i+m}(k)+\frac{\delta_{ij}}{U}, \qquad \qquad  \qquad\qquad (1\leq i,j\leq m)\notag\\
 &&M_{ij}(\textbf{q},i\omega_n)=\frac{1}{\beta}\sum_{\textbf{k},n'}G^{0}_{ij}(k+q)^{0}G_{j-m,i+m}(k), \qquad \qquad \qquad \qquad (1\leq i\leq m \;\&\; m+1\leq j\leq 2m)\notag\\
&&M_{ij}(\textbf{q},i\omega_n)=\frac{1}{\beta}\sum_{\textbf{k},n'}G^{0}_{ij}(k+q)G^{0}_{j+m,i-m}(k), \qquad  \qquad \qquad \qquad(m+1\leq i\leq 2m \; \& \; 1\leq j\leq m)\notag\\
 &&M_{ij}(\textbf{q},i\omega_n)=\frac{1}{\beta}\sum_{\textbf{k},n'}G^{0}_{ij}(k+q)G^{0}_{j-m,i-m}(k)+\frac{\delta_{ij}}{U}, \qquad \qquad \qquad \qquad(m+1\leq i,j\leq 2m ),
\end{eqnarray}
where the number of order parameters $m=3$ and Green function  $G^{0}_{ij}(\textbf{k},i\omega_{n'})= ([i\omega_{n'}-H_{\rm BdG}(\textbf{k})]^{-1})_{ij}$
is matrix element of Nambu-Gorkov Green function. The collective modes are given by zeroes of determinant ${\rm Det}|M(\textbf{q},i\omega_n\rightarrow \omega+i0^+)|=0$.

\bibliography{sample}



\section*{Acknowledgements (not compulsory)}
 Y. R. W. and Y.C. Z. are supported by the NSFC under Grants No.11874127 and startup grant from Guangzhou University. X.F.Z. is Supported by the National Natural Science Foundation of
China (Grant No. 11775253) and the Key Research Program of Frontier Sciences, CAS under Grant No. ZDBS-LY-7016.
C. F. L. is supported by the NSFC under grant Nos. 11875149 and 61565007, The Youth Jinggang Scholars Program in Jiangxi Province, and The Program of Qingjiang Excellent Yong Talents, Jiangxi University of Science and Technology. W.M. L. is supported by the National Key R\&D Program of China under grants No. 2016YFA0301500, NSFC under grants Nos.11434015, 61227902.

\section*{Author contributions statement}
 Y.R W. and Y.C.Z. performed calculations. Y.R W., X.F.Z., C.F.L., W.M.L., and  Y.C.Z. analyzed numerical results. Y.R W., X.F.Z., C.F.L., W.M.L., and  Y.C.Z. contributed in completing the paper.

\section*{Additional information}

Competing interests: The authors declare no competing interests

The corresponding author is responsible for submitting a \href{http://www.nature.com/srep/policies/index.html#competing}{competing interests statement} on behalf of all authors of the paper. 




\end{document}